\documentclass[twocolumn,preprintnumbers,superscriptaddress,nofootinbib,aps,prd,floatfix]{revtex4}

\usepackage{enumerate}
\usepackage{amsmath,amssymb}
\usepackage{graphicx}
\usepackage{bbm,slashed}
\usepackage{xspace,slashed}
\usepackage{hyperref}
\hypersetup{colorlinks=true, citecolor=blue, urlcolor=blue, linkcolor=blue}
\usepackage[normalem]{ulem}

\newcommand{\mg}{\textsc{MadGraph5\_}aMC@NLO}
\newcommand{\pythia}{\textsc{Pythia8}}

\begin{document}

\providecommand{\abs}[1]{\lvert#1\rvert}

\newcommand{\Znunujets}{(Z\to{\nu\bar{\nu}})+\text{jets}}
\newcommand{\Welnujets}{(W\to{\ell\nu})+\text{jets}}
\newcommand{\Znunujet}{(Z\to{\nu\bar{\nu}})+\text{jet}}
\newcommand{\Welnujet}{(W\to{\ell\nu})+\text{jet}}

\title{Machine-enhanced CP-asymmetries in the Higgs sector}
\begin{abstract}
Improving the sensitivity to CP-violation in the Higgs sector is  one of the pillars of the precision Higgs programme at the Large Hadron Collider. We present a simple method that allows CP-sensitive observables to be directly constructed from the output of neural networks. We show that these observables have improved sensitivity to CP-violating effects in the production and decay of the Higgs boson, when compared to the use of traditional angular observables alone. The kinematic correlations identified by the neural networks can be used to design new analyses based on angular observables, with a similar improvement in sensitivity. 
\end{abstract}

\author{Akanksha Bhardwaj}\email{akanksha.bhardwaj@glasgow.ac.uk} 
\affiliation{School of Physics \& Astronomy, University of Glasgow, Glasgow G12 8QQ, UK\\[0.1cm]}
\author{Christoph Englert} \email{christoph.englert@glasgow.ac.uk}
\affiliation{School of Physics \& Astronomy, University of Glasgow, Glasgow G12 8QQ, UK\\[0.1cm]}
\author{Robert Hankache}\email{robert.hankache@manchester.ac.uk} 
\affiliation{Department of Physics \& Astronomy, University of Manchester, Manchester M13 9PL, UK\\[0.1cm]}
\author{Andrew D. Pilkington}\email{andrew.pilkington@manchester.ac.uk} 
\affiliation{Department of Physics \& Astronomy, University of Manchester, Manchester M13 9PL, UK\\[0.1cm]}


\maketitle

\section{Introduction}
\label{sec:intro}
The Sakharov criteria~\cite{Sakharov:1967dj} provide the theoretical backdrop for one of the biggest phenomenological shortfalls of the Standard Model (SM) of Particle Physics -- an insufficient amount of charge-conjugation (C) and parity~(P) violation. In the SM, the only source of CP violation is the complex phase in the Cabibbo-Kobayashi-Maskawa~(CKM) matrix~\cite{Cabibbo:1963yz,Kobayashi:1973fv}. As the flavour and CP structure of SM interactions is intricately related to the Yukawka sector, extending the Higgs sector with additional CP-violating effects is typically considered as a motivated avenue to reconcile the SM with the Sakharov criteria.

Such extensions of the SM typically lead to new exotic states~\cite{DasBakshi:2021iey}, which so far have not been discovered at the Large Hadron Collider (LHC). This suggests that there is a significant gap between the mass scale of weak interactions and the mass scale of beyond-the-SM (BSM) physics. This line of thought has led to a resurgence of effective field theory applications to the interpretation of LHC data~\cite{Buchmuller:1985jz,Burges:1983zg,Leung:1984ni,Hagiwara:1986vm,Grzadkowski:2010es,Jenkins:2013wua,Alonso:2013hga,Jenkins:2013zja,Elias-Miro:2014eia,Brivio:2017btx,deBlas:2017xtg,Helset:2018fgq}. The extension of the SM by dimension-six interactions provides the first step in this programme, capturing the deformations of correlations in particle physics data under the assumption that there is a hierarchy between the scale of measurement and new physics $Q^2\ll \Lambda^2$. Of particular interest are the operators, $\widetilde{{\cal{O}}}_i$, that introduce new sources of CP violation in the Lagrangian, 
\begin{equation}
\label{eq:eff}
    {\cal{L}}={\cal{L}}_{\text{SM}} + \sum\limits_{i} {c_i\over \Lambda^2} \widetilde{{\cal{O}}}_i\,,
\end{equation}
where ${\cal{L}}_{\text{SM}}$ is the SM Lagrangian and the ${c_i / \Lambda^2}$ are Wilson coefficients that specify the strength of the new interactions. The operators that affect the electroweak interactions of the Higgs boson are~(see also~\cite{Degrande:2021zpv})
\allowdisplaybreaks
\begin{equation}
\label{eq:ops1}
\begin{split}
    {\cal{O}}_{\Phi \widetilde{B}} &= \Phi^\dagger \Phi  B^{\mu\nu}\widetilde{B}_{\mu\nu}\,,\\
  {\cal{O}}_{\Phi \widetilde{W}} &=  \Phi^\dagger \Phi  W^{i\,\mu\nu}\widetilde{W}^{i}_{\mu\nu}\,, \\
  {\cal{O}}_{\Phi \widetilde{W}B} &=  \Phi^\dagger \sigma^i  \widetilde{W}^{i\,\mu\nu}B_{\mu\nu}\,, 
\end{split}
\end{equation}
where $\Phi$ is the Higgs field, 
and the $W^\mu$ and $B^\mu$ are the fields in the $SU(2)\otimes U(1)$ gauge-field eigenbasis. The dual field strength tensors are defined as $\widetilde X^{\mu\nu}=\epsilon^{\mu\nu\rho\delta}X_{\rho\delta}/2$.\footnote{Additionally, phases of Wilson coefficients can introduce CP violation in the Higgs-fermion interactions.}

The contributions of these operators to Higgs boson production and decay is given by the squared amplitude, i.e.
\begin{multline}
\label{eq:xsec}
|{\cal{M}}|^2 = |{\cal{M}}_{\text{SM}}|^2 \\
    + {c_i\over \Lambda^2}\, 2\, \Re \left[ {\cal{M}}_{\text{SM}}{\cal{M}}_{\text{d6}, i}^\ast \right] +  {c_i\, c_j\over \Lambda^4} {\cal{M}}_{\text{d6},i}{\cal{M}}_{\text{d6},j}^\ast\,,
\end{multline}
where ${\cal{M}}_{\text{SM}}$ and ${\cal{M}}_{\text{d6},i}$ are the SM and dimension-six amplitudes, respectively. For the CP-odd operators of interest, the interference between the SM amplitude and the dimension-six amplitude is also CP-odd. Interference effects therefore cancel entirely for CP-even observables, such as inclusive cross sections and transverse-momentum spectra, but can be observed as asymmetries in appropriately-constructed CP-odd observables~\hbox{\cite{Buszello:2002uu,Choi:2002jk,Buszello:2004be,Godbole:2007cn,Gao:2010qx,DeRujula:2010ys,Englert:2010ud,Bernlochner:2018opw,Plehn:2001nj,Hankele:2006ma,Klamke:2007cu,Campanario:2010mi,Englert:2012ct,Anderson:2013afp,Campanario:2013mga,Campanario:2014oua,Chen:2014ona,Buckley:2015vsa,Brehmer:2017lrt,Goncalves:2018agy,Davis:2021tiv,Banerjee:2020vtm}}. The inclusion of the pure dimension-six contributions to the amplitude-squared in Eq.~\eqref{eq:xsec} gives two potential problems. First, these contributions are CP-even, making it difficult to disentangle the effects of a CP-even operator from a CP-odd operator. Second, the contributions arise at ${\cal{O}}(1/\Lambda^4)$ and power counting of the new physics scenario becomes important in this instance, i.e. it is a model-dependent question whether the leading ${\cal{O}}(1/\Lambda^2)$ dominate over the ${\cal{O}}(1/\Lambda^4$) expansion in an actual matching calculation~\cite{DasBakshi:2020ejz,DasBakshi:2021iey}. 

For these reasons, the ATLAS and CMS experiments have an extensive programme of searches and measurements that utilise CP-odd observables, including angular observables~\cite{ATLAS:2018hxb,ATLAS:2020wny,ATLAS:2021pkb,CMS:2021sdq} as well as observables that are constructed from matrix-element information~\cite{ATLAS:2016ifi,CMS:2019jdw,ATLAS:2020evk,CMS:2021nnc}. The latter approach exploits the full kinematic information in leading-order matrix-elements to discriminate different CP hypotheses, and is shown to improve the analysis sensitivity over the use of angular observables alone. The use of matrix-elements in an analysis is, however, more technically challenging and time-consuming than using the simpler angular observables. For this reason, only a few experimental analyses have adopted these more sophisticated analysis techniques to date.\footnote{Recently, it was shown that machine-learning algorithms can be used to construct a discriminant that is equivalent to the discriminant constructed from matrix-element information, with a slight degradation in ROC-curve performance that is attributed to imperfect training~\cite{Gritsan:2020pib}. This approach could help overcome the time-consuming aspects of using matrix-element-based observables in physics analysis.}

In this article, we show that CP-odd observables can be directly constructed from the output of a neural network. Given that the ${\cal{O}}(1/\Lambda^2$) interference effects cancel entirely for CP-even observables, and induce asymmetries in CP-odd observables, we can directly construct a CP-odd observable by training a neural network~(NN) to distinguish between positive and negative interference contributions. With the ability to learn kinematic correlations, the NN can be used to (i) construct a near-optimal CP-odd observable for each dimension-six operator, or (ii) design new  analyses based on the correlation between the angular observables and other kinematic quantities. The method can then be extended to multi-class models, with the pure-SM prediction included in the training of the network, to allow the NN to learn the phase-space regions for which the SM is suppressed relative to the interference contribution. 

As a concrete example, we explore the potential application of neural networks in two of the main search channels for CP-violation in the Higgs sector: the $h\to 4\ell$ decay channel and in the vector-boson fusion production channel (VBF $h+2~\text{jets}$). As well as addressing the phenomenological difference between Higgs production and Higgs decay, the comparison of $h\to 4\ell$ and $h+2~\text{jets}$ also highlights the difference between single-scale and multi-scale processes when viewed through a NN lens. We note that the technique should be applicable to a wide variety of production and decay channels at the LHC (see also the recent \cite{Ren:2019xhp,Bortolato:2020zcg,Barman:2021yfh}).

We organise the work as follows. In Section~\ref{sec:mc}, we introduce the Monte Carlo event generators that we use to construct the SM and dimension-six theoretical predictions for Higgs boson production at the LHC. In Section~\ref{sec:cp}, we recap the angular observables that typically are used for CP-violation searches in the $h\to 4\ell$ and $h+2~\text{jets}$ final states. We also introduce the method to construct CP-odd observables using neural networks. In Section~\ref{sec:results}, we apply this method to simulated $h\to 4\ell$ and $h+2~\text{jets}$ events and compare the sensitivity of the machine-learned CP-odd observables to the sensitivity obtained using angular observables. We also investigate the origin of any improvement in sensitivity. Finally, we conclude in Section~\ref{sec:conc}. 

\section {Theoretical predictions}
\label{sec:mc}
Events are generated for the production of $h\to 4\ell$ and VBF $h+2\rm{jet}$ in proton-proton collisions at $\sqrt{s}=13$~TeV using \mg{}~\cite{Alwall:2014hca}. The events are accurate to leading order in perturbative QCD and are passed to \pythia{}~\cite{Sjostrand:2007gs} to simulate the effects of parton-showering, hadronisation and underlying event activity. The NNPDF30nlo (NNPDF23lo) parton distribution function~\cite{Ball:2012cx} is used in the cross-section calculation for the $h\to 4\ell$  ($h+2\rm{jet}$) samples. The A14 set of tuned parameters~\cite{ATL-PHYS-PUB-2014-021} is used to model the underlying event. Events are generated separately for the Standard Model and for the interference between the SM and dimension-six amplitudes, with the interactions induced by the dimension-six operators provided by the \hbox{SMEFTSim} package~\cite{Brivio:2017btx}. In the $h+2~\rm{jets}$ sample, the Higgs boson is not decayed, as we focus on production-related kinematics in this channel.

For the $h \to 4 \ell{}$ analysis, we require the generated events to pass the selection criteria of the ATLAS $pp\to 4\ell$ measurement~\cite{ATLAS:2021kog}, in the \textit{Higgs Mass} fiducial region ($120~\text{GeV} < m_{4\ell}< 130$~GeV). For the analysis of VBF Higgs production, we require the events to pass the selection criteria of the ATLAS VBF $h\to \tau^+\tau^-$ analysis~\cite{ATLAS-CONF-2021-044}, in the \textit{VBF\_1} fiducial region.

\section {CP-sensitive observables}
\label{sec:cp}
\subsection{Angular observables}

CP-violating effects in the $h\to ZZ^*\to 4\ell$ decay channel can be probed using the $\Phi_{4\ell{}}$ variable \cite{Bolognesi:2012mm,Gritsan:2016hjl} defined by
\begin{equation}
\label{eq:cpodd}
    \Phi_{4\ell{}} = \frac{{\bf q}_1 \cdot ( \hat{\bf n}_1 \times \hat{\bf n}_2)}{| {\bf q}_1 \cdot \left( \hat{\bf n}_1 \times \hat{\bf n}_2\right)}| \times {\cos}^{-1}({\bf \hat{\bf n}_1 \cdot \hat{\bf n}_2)},
\end{equation}
where the normal vectors to the planes are defined as
\begin{equation}
    {\hat{\bf n}_1} = \frac{{\bf q}_{11} \times {\bf q}_{12}}{| {\bf q}_{11} \times {\bf q}_{12}|} \quad {\text{and}} \quad \hat{\bf n}_2 = \frac{{\bf q}_{21} \times {\bf q}_{22}}{|{\bf q}_{21} \times {\bf q}_{22}|}.
\end{equation}
Each ${\bf q}_{\alpha\beta}$ labels the three-momentum of the lepton/antilepton $\beta$ that arises from the decay $Z_\alpha \to \ell \bar{\ell}$, and ${\bf q}_\alpha = {\bf q}_{\alpha1} + {\bf q}_{\alpha2}$ is the three momentum of the $Z_\alpha$. All three-momenta are calculated in the Higgs-boson centre-of-mass frame. It is worth noting that $\Phi_{4\ell{}}$ coincides with the angular difference of the polar angles of the leptons with identical charge in their respective $Z$ boson rest frame (for aligned reference axes).

CP-violating effects in the VBF $h + 2~{\text{jets}}$ production channel can be probed using the signed azimuthal angle between the two jets, i.e. 
\begin{equation}
    \label{eq:signedphijj}
    \Delta\phi_{jj} = \phi(j_1) - \phi(j_2)\,,\quad \text{with}~y(j_1)>y(j_2)\,,
\end{equation}
where $\phi(j_1)$ and $\phi(j_2)$ are the azimuthal angles of the two highest transverse momentum jets in the event that are ordered in rapidity $y$. 
The interference effects and associated asymmetry effects can be traced to the vertex structure that is induced by the operators of  Eq.~\eqref{eq:ops1}. The Levi-Civita tensor determines a $P$-odd behaviour of the decay amplitude
\begin{equation}
  \label{eq:d6ep}
    {\cal{M}}_{d6} \sim \epsilon_{\mu\nu\rho\delta}\, j_1^\mu(q_1)j_2^\nu(q_2) q^\rho_{1}q^\delta_{2} 
\end{equation}
where $q_i$ are the two four momenta of the effective fermion currents $j_i$ coupling to the Higgs boson. C and P transformations induce sign changes of the currents $j^\mu_i,q^\mu_i$. Together with the odd property of the Levi-Civita tensor under parity transformations, this leads to an asymmetry of the interference effects as a function of $\Delta\phi_{jj}$ (see also~\cite{Plehn:2001nj}). Note also that ${\cal{M}}_{d6}$ is only non-vanishing for linear independent momenta and effective currents, thus removing longitudinal effective polarisations from the BSM amplitude. In the case of VBF production, the currents in Eq.~\eqref{eq:d6ep} are related to interactions between the tagging jet and its associated initial state parton. For VBF kinematics $p_{T,j}\ll E_j$ this leads to a $p_T$-enhanced CP-sensitivity (the aforementioned optimal observable), which is reflected in the VBF results below.

\subsection{ML-constructed CP-odd observables}
We use {\sc{TensorFlow 2.3.0}} \cite{Abadi:2016kic} to train the neural networks. The input data are the MC samples discussed in Sec.~\ref{sec:mc}, with the events in the interference sample separated according to whether the event weight was positive or negative. Two types of neural network architectures are investigated. {\emph{Binary}} (two-class) models are trained using only the interference sample and define the probability that a given event is a positively-weighted interference event ($P_+$) or a negatively-weighted interference event ($P_-$). In these models, $P_+ + P_- = 1$. \textit{Multi-class} models are trained using both the interference sample and the pure-SM prediction, and therefore also define the probability that a given event is a SM event ($P_{\text{SM}}$). In these models, $P_+ + P_- + P_{\text{SM}} = 1$. The machine-learned CP-odd observable is then defined by
\begin{equation}
    O_{NN} = P_+ \, - \, P_- \, .
\end{equation}

The ability of a neural network to construct the CP-odd observable is, in principle, dependent on the input information, with the simplest input being only the four-vectors of the leptons and jets. More complex inputs would include variables that can be derived from those four-vectors, such as $\Phi_{4\ell{}}$  in the $h\to 4\ell{}$ decay channel. In general, we find that the neural networks perform equally well when including derived variables or just using lepton and jet four-vectors. However, the inclusion of derived variables can help with understanding the physical origin of any improvement in sensitivity. Unless otherwise stated, the results presented in this article use neural networks trained with both lepton/jet four vectors and derived variables.

The optimal choice of hyperparameters for each network is obtained using {\sc{Keras-Tuner 1.0.2}} \cite{keras,omalley2019kerastuner}. The optimisation included the number of layers, the number of units, the activation function, the L2 regularisation, the learning rate and the batch size. 
To avoid the networks exploiting statistical fluctuations, we adopt a data augmentation procedure whereby each event is used twice in the training, once with the default input variables and once with a CP-operator applied to all the input variables. For CP-flipped events in the interference sample, the event weight is multiplied by -1. After the initial training, in order to smooth the model, we train for more epochs using the full batch. We also apply a learning rate decay, beginning with the initial rate and halving it every 100 epochs until reaching a factor of 1/8.

\section{Results}
\label{sec:results}
\subsection{$h\to 4\ell$}
\label{sec:h4ell}
The construction of $\Phi_{4\ell{}}$ and $O_{NN}$ requires each lepton and antilepton to be associated with the decay of an intermediate $Z$-boson.
For the $h\to e^+e^- \mu^+ \mu^-$ decay channel, this is trivial because the $Z$-boson always decays to a same-flavour opposite-charge pair. However, an ambiguity arises in the $h\to e^+ e^- e^+ e^-$ and $h\to \mu^+\mu^- \mu^+ \mu^-$ decay channels, due to the multiple possible pairings of the leptons and antileptons. For this reason, we initially restrict our discussion to the $h\to e^+e^- \mu^+ \mu^-$ decay channel and comment later on the performance of the other $h\to 4\ell$ decay channels. 

\begin{figure}[t!]
\centering
    \includegraphics[width=0.45\textwidth]{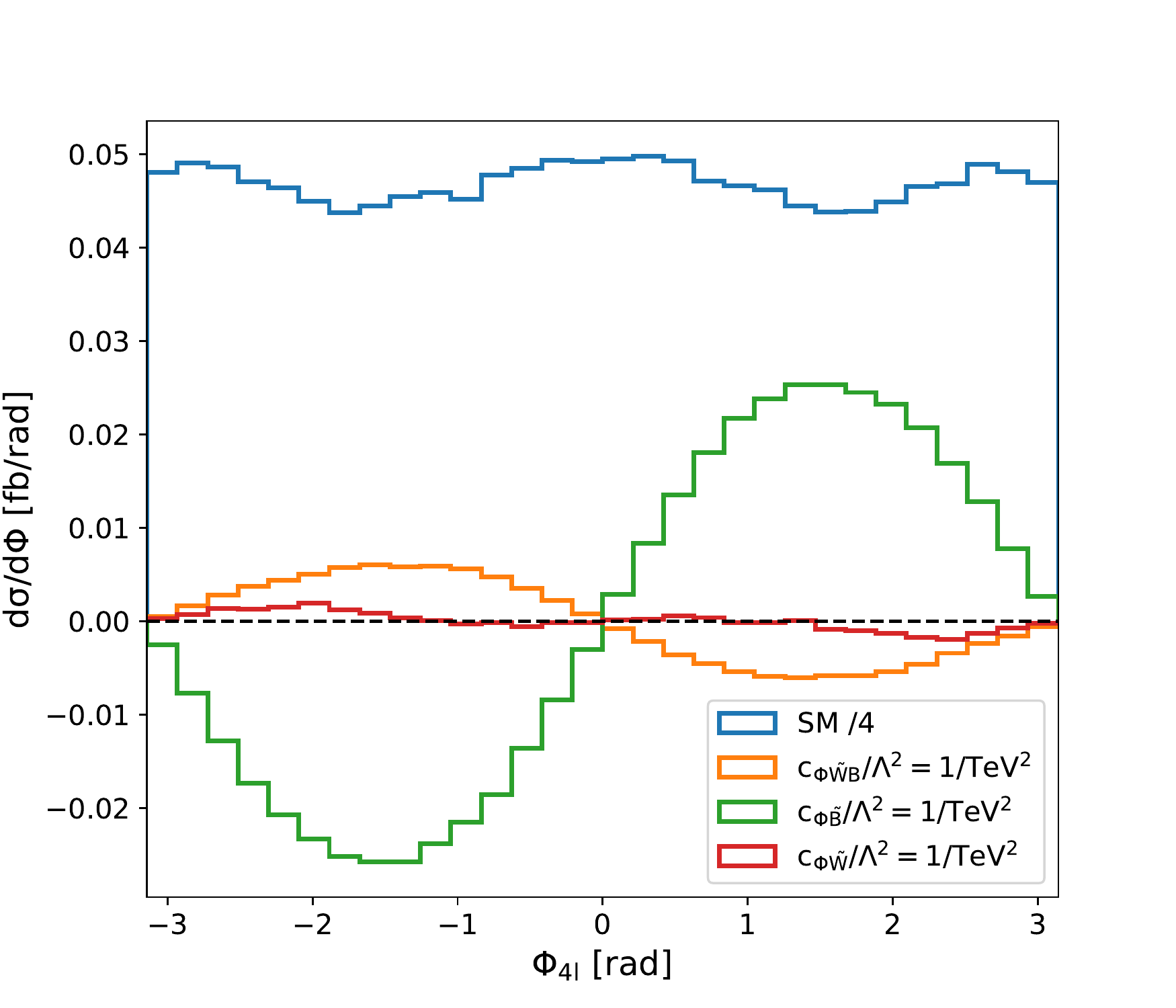}
  \caption{Differential cross section for $h\to e^+e^-\mu^+\mu^-$ as a function of $\Phi_{4\ell{}}$. The interference predictions obtained for the ${\cal{O}}_{\Phi \widetilde{B}}$, ${\cal{O}}_{\Phi \widetilde{W}B}$ and ${\cal{O}}_{\Phi \widetilde{W}}$ operators are shown, with Wilson coefficients set to $c/\Lambda^2=1$~TeV$^{-2}$. The SM prediction is also shown, scaled down by a factor of 4.}
  \label{fig:phi4l_xs}
\end{figure}

The differential cross section for $h\to e^+e^-\mu^+\mu^-$ as a function of $\Phi_{4\ell{}}$ is presented in Fig.~\ref{fig:phi4l_xs}. The SM prediction is shown in addition to the interference contributions induced by the ${\cal{O}}_{\Phi \widetilde{B}}$, ${\cal{O}}_{\Phi \widetilde{W}B}$ and ${\cal{O}}_{\Phi \widetilde{W}}$ operators,   with Wilson coefficients set to $c/\Lambda^2=1$~TeV$^{-2}$. As expected, the CP-even SM prediction is symmetric around $\Phi_{4\ell{}}=0$, whereas the CP-odd interference contributions are all asymmetric with an integral of zero. The largest interference effects arise from the ${\cal{O}}_{\Phi \widetilde{B}}$ operator. The $\Phi_{4\ell{}}$ distribution is much less sensitive to the ${\cal{O}}_{\Phi \widetilde{W}B}$ and ${\cal{O}}_{\Phi \widetilde{W}}$ operators, and much larger values of Wilson coefficient would be needed to produce a noticeable effect on the combined SM$+$EFT cross section.

\begin{figure}[t!]
\centering
    \includegraphics[width=0.45\textwidth]{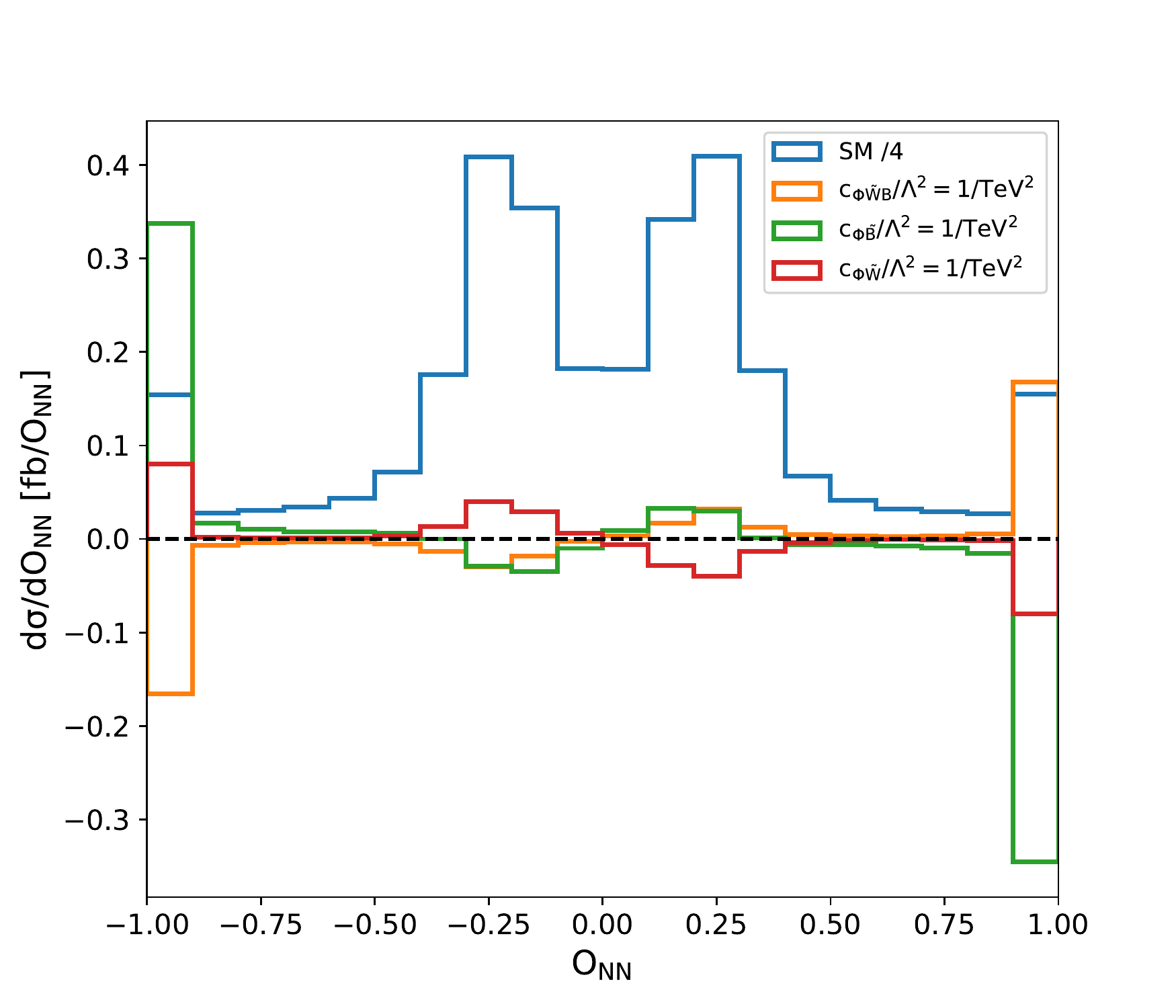}
  \caption{Differential cross section as a function of the CP-odd observable, $O_{NN}$, constructed for a binary neural network. The network was trained with the interference predictions obtained with the ${\cal{O}}_{\Phi \widetilde{W}B}$ operator. The interference predictions obtained for the ${\cal{O}}_{\Phi \widetilde{B}}$, ${\cal{O}}_{\Phi \widetilde{W}B}$ and ${\cal{O}}_{\Phi \widetilde{W}}$ operators are shown, with Wilson coefficients set to $c_{\Phi \widetilde{W}B}/\Lambda^2=1$~TeV$^{-2}$. The SM prediction is also shown, scaled down by a factor of~4.}
  \label{fig:NNSC_xs}
\end{figure}

The differential cross section as a function of the CP-odd observable produced by a binary NN is shown in Fig.~\ref{fig:NNSC_xs}, where the NN has been trained to distinguish between the positive- and negative- interference effects produced by the ${\cal{O}}_{\Phi \widetilde{W}B}$ operator.  The interference contribution is presented for $c_{\Phi \widetilde{W}B}/\Lambda^2=1$~TeV$^{-2}$. The NN effectively separates the positively-weighted and negatively-weighted interference contributions, with the majority of positively-weighted events located at $O_{NN}=1$ and the majority of the negatively-weighted events located at $O_{NN}=-1$. The SM contribution is symmetric and more broadly distributed, peaking at NN output values closer to zero. Most importantly, the SM contribution in the interference-enhanced regions at $O_{NN}=\pm1$ is reduced when compared to the interference-enhanced regions in $\Phi_{4\ell{}}$, implying a possible improvement in sensitivity due to the increased signal purity. This is especially noticeable for the ${\cal{O}}_{\Phi \widetilde{W}B}$ and ${\cal{O}}_{\Phi \widetilde{W}}$ operators.

\begin{figure}[t!]
\centering
    \includegraphics[width=0.45\textwidth]{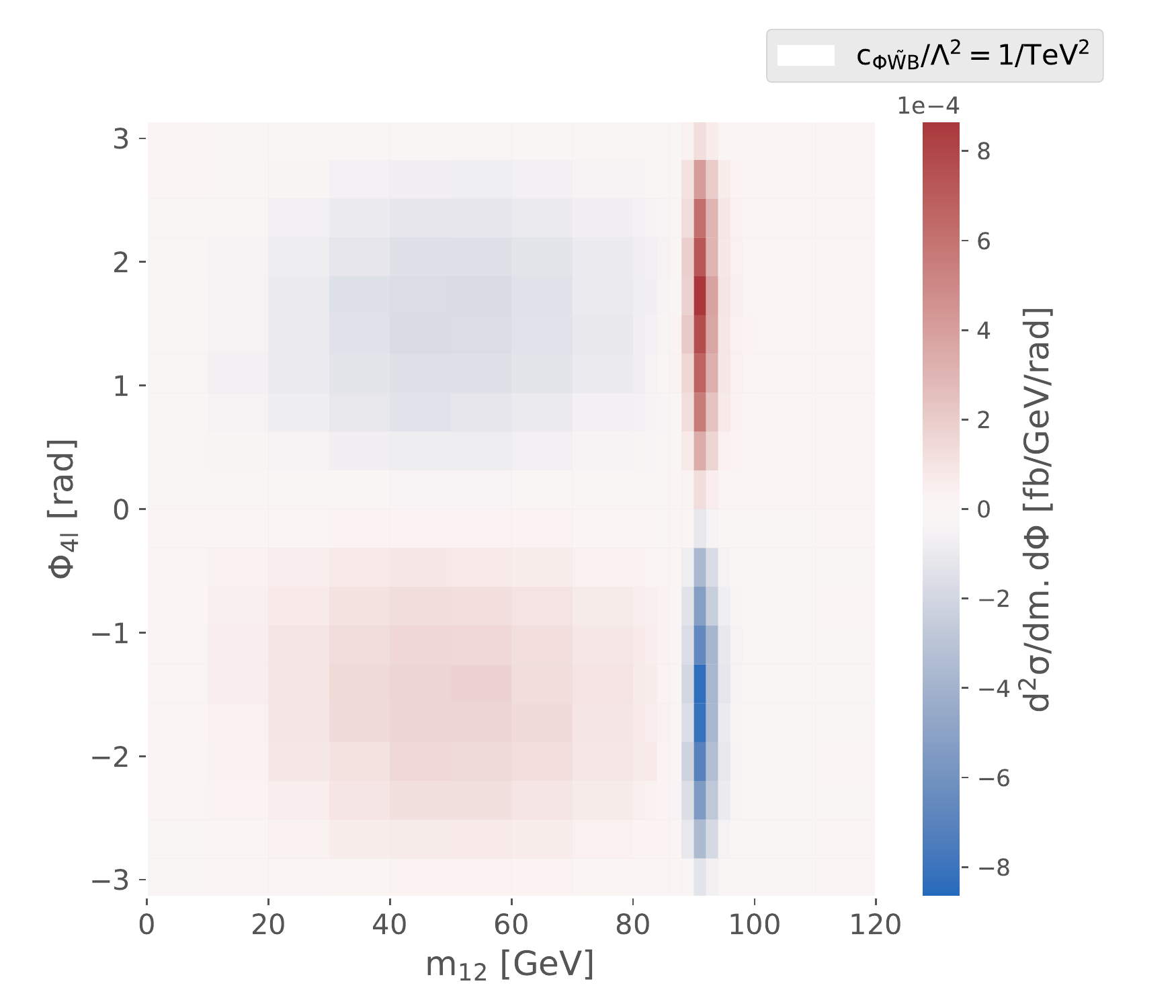}
  \caption{Double-differential cross section as a function of $\Phi_{4\ell{}}$ and $m_{12}$ for the interference contribution produced by the ${\cal{O}}_{\Phi \widetilde{W}B}$ operator.}
  \label{fig:m4lphi4l}
\end{figure}

The improved sensitivity obtained using the neural network can be understood using feature importance techniques. Specifically, the importance of each input variable is determined for the trained network, by evaluating the increase in the loss (or decrease in the accuracy) that occurs when the value of the input variable for a given event is replaced by a randomly chosen value taken from the ensemble of events. 
Unsurprisingly, the most-important variable is found to be $\Phi_{4\ell{}}$. However, the invariant mass ($m_{12}$) of the lepton-antilepton pair that is closest in mass to the $Z$-boson is also found to be very important, despite being a CP-even quantity. This is explored more in more detail in Fig.~\ref{fig:m4lphi4l}, which shows the double-differential cross section for the interference contribution induced by the ${\cal{O}}_{\Phi \widetilde{W}B}$ operator as a function of $\Phi_{4\ell{}}$ and $m_{12}$. The importance of $m_{12}$ is clear: at a given value of $\Phi_{4\ell{}}$, the interference effects for events with $m_{12}\sim m_Z$ are opposite in sign to the interference effects for events with $m_{12}\ll m_Z$, which cancel when $\Phi_{4\ell{}}$ is measured inclusively. The neural network has learned this feature and utilised it to produce an improved CP-odd observable.

The origin of the sign flip in the interference contribution induced by the ${\cal{O}}_{\Phi \widetilde{W}B}$ operator is driven by the anomalous $hZZ$, $h\gamma\gamma$ and $hZ\gamma$ interactions, which are related to one another via gauge symmetry and are given by
\begin{equation}
\begin{split}
 C_{H\widetilde{Z}Z} &=    c_w^2C_{H\widetilde{W}} + s_w (c_wC_{H\widetilde{W}B} +s_w C_{H\widetilde{B}} ) \\
 C_{H\widetilde{A}A} &=  c_w^2 C_{H\widetilde{B}}+ s_w (s_w C_{H\widetilde{W}}  - c_wC_{H\widetilde{W}B} )\,,\\
 C_{H\widetilde{A}Z} &= 2s_w c_w (  C_{H\widetilde{W}}-C_{H\widetilde{B}} ) +(s_w^2 - c_w^2) C_{H\widetilde{W}B}\,,
\end{split}
\end{equation}
where $s_w$ and $c_w$ are the sine and cosine of the Weinberg angle, respectively. The impact of the ${\cal{O}}_{H\widetilde{W}B}$ operator is anticorrelated for the $hZZ$ and $hZ\gamma$ ($h\gamma\gamma$) anomalous interactions, which leads to anticorrelated interference contributions. The sign flip therefore occurs due to different contributions from the $h\to ZZ$, $h\to \gamma\gamma$ and $h \to Z\gamma$ dimension-six amplitudes in the on-shell and off-shell regions.

\begin{figure}[t!]
\centering
    \includegraphics[width=0.45\textwidth]{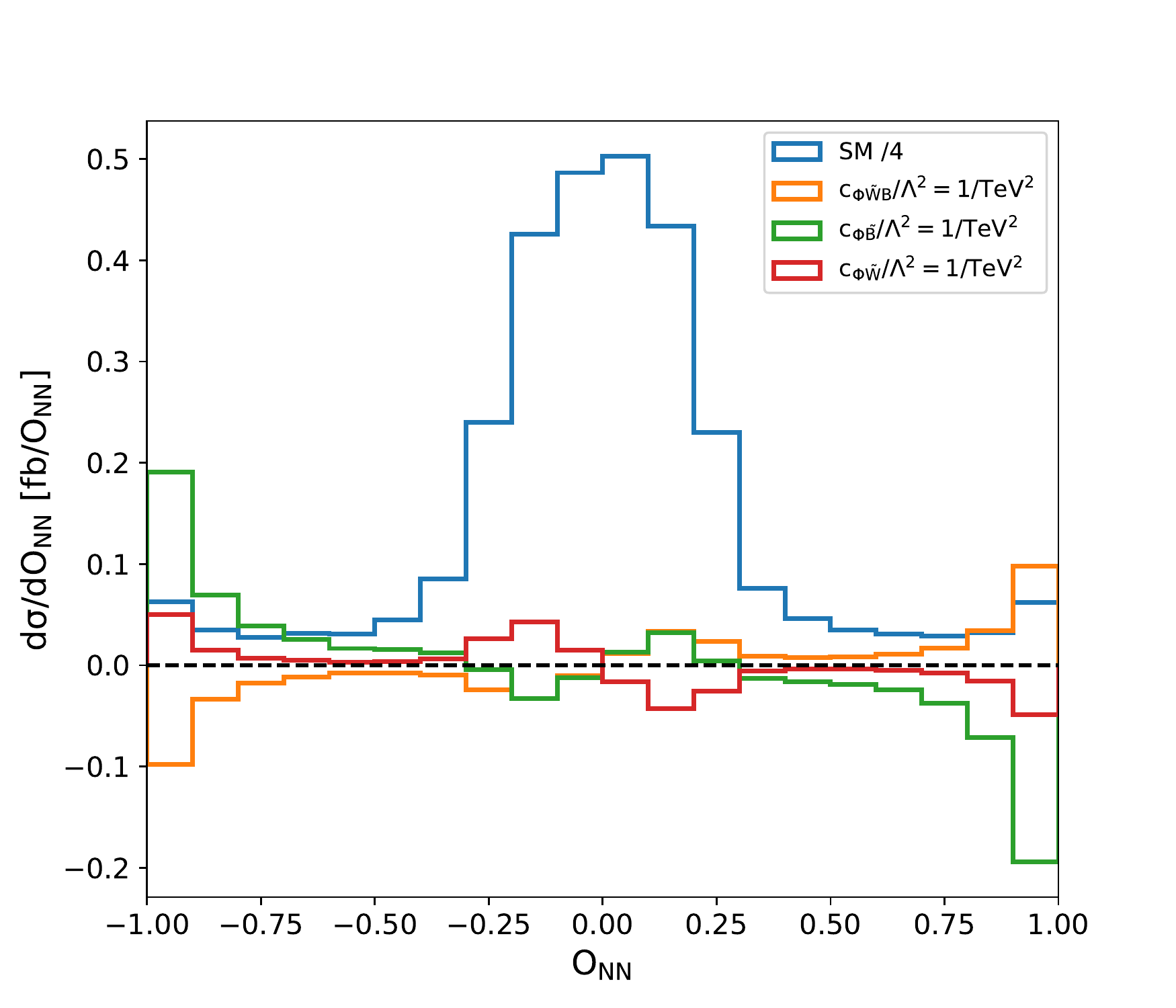}
  \caption{Differential cross section as a function of the CP-odd observable, $O_{NN}$, constructed for a multi-class neural network. The network was trained with the interference predictions obtained with the ${\cal{O}}_{\Phi \widetilde{W}B}$ operator. The interference predictions obtained for the ${\cal{O}}_{\Phi \widetilde{B}}$, ${\cal{O}}_{\Phi \widetilde{W}B}$ and ${\cal{O}}_{\Phi \widetilde{W}}$ operators are shown, with Wilson coefficients set to $c_{\Phi \widetilde{W}B}/\Lambda^2=1$~TeV$^{-2}$. The SM prediction is also shown, scaled down by a factor of 4.}
  \label{fig:NNMC_xs}
\end{figure}

The sensitivity of $O_{NN}$ can be further improved by using multi-class neural networks, which have the ability to learn the kinematic features of the SM prediction. Figure~\ref{fig:NNMC_xs} shows the differential cross section as a function of the CP-odd observable constructed from a multi-class network. The neural network has been trained to distinguish between the SM contribution as well as the positive- and negative- interference effects produced by the ${\cal{O}}_{\Phi \widetilde{W}B}$ operator. The interference contributions are still peaked at $O_{NN}=\pm1$, but with a broader peak than what was obtained with a binary neural network. However, the SM prediction is shifted much closer to (and peaks at) zero. Overall, the SM contribution in the interference-enhanced regions at $O_{NN}=\pm1$ is further reduced when compared to the binary network, implying a further increase in sensitivity.

To quantify the sensitivity of an experimental analysis, we construct the expected 95\% confidence intervals for each CP-odd observable. A likelihood function is defined as
\begin{align}
 L_i(\boldsymbol{c}/\Lambda^2) =  \prod_{i} \frac{{e}^{-\lambda_i}}{n_i!} \lambda_i^{n_i} 
\end{align} 
where $n_i$ is the expected number of events in bin $i$ assuming the SM-only hypothesis, and $\lambda_i$ is the predicted number of events (SM$+$EFT) at a given value of a Wilson coefficient. The expected number of events are obtained from the event generator samples after applying a normalisation factor that is defined such that the SM prediction reproduces the number of events observed experimentally in the \textit{Higgs mass} fiducial region of Ref.~\cite{ATLAS:2021kog}, corresponding to an integrated luminosity of 139~fb$^{-1}$.
The confidence level is then calculated using the profile-likelihood test statistic~\cite{Feldman:1997qc}, which is assumed to be distributed according to a $\chi^2$ distribution with one degree of freedom following from Wilks' theorem~\cite{wilks1938} and allows the 95\% confidence intervals to be constructed.\footnote{This assumption is validated by constructing pseudo-experiments to determine the distribution of the profile-likelihood test statistic. The resulting distribution is well modelled by a $\chi^2$ distribution with one degree of freedom.}  Although the likelihood function does not account for systematic uncertainties, the effect of any systematic variation should be symmetric for all CP-odd observables. As the constraints are driven by asymmetries in the distribution, the impact of systematic uncertainties should be very small. This was tested by injecting small symmetric shifts into the predicted number of events, to simulate a systematic bias. The resulting 95\% confidence intervals were almost unchanged.

The constraints obtained for each Wilson coefficient are shown in Tab.~\ref{tab:4lsummary}, when performing a fit to (i) the angular observable $\Phi_{4\ell{}}$, (ii) a two-dimensional fit to $\Phi_{4\ell{}}$ and $m_{12}$, and (iii) fits to the NN-constructed $O_{NN}$ observables for both binary and multi-class networks. The $O_{NN}$ observables were obtained with networks trained on the interference predictions obtained with the ${\cal{O}}_{\Phi \widetilde{W}B}$ operator. The $O_{NN}$ observables both provide much better sensitivity than $\Phi_{4\ell{}}$ alone, with 95\% confidence intervals reduced by a factor of 2-10, depending on the Wilson coefficient. Some of this improvement is regained using a two-dimensional fit to $\Phi_{4\ell{}}$ and $m_{12}$, although the constraints obtained using the $O_{NN}$ observables remain 20-30\% more sensitive. The constraints obtained from the CP-odd observable constructed from the multi-class network are 5-10\% better than the constraints obtained from using binary networks. It is also found that the constraints on $c_{\Phi \widetilde{B}}$  and $c_{\Phi \widetilde{W}}$ can be further improved by 5\% and 10\%, respectively, if the neutral network is specifically trained on the interference predicted by the associated operators.

\begin{table*}[t!]
    \centering
    \begin{tabular}{|c|c|c|c|}
    \hline 
     CP-odd observable & $c_{\Phi \widetilde{W}B} / \Lambda^2$~[TeV$^{-2}$] & $c_{\Phi \widetilde{B}} / \Lambda^2$~[TeV$^{-2}$] &  $c_{\Phi \widetilde{W}} / \Lambda^2$~[TeV$^{-2}$] \\
    \hline 
       $\Phi_{4\ell{}}$  & [-6.2,6.2] & [-1.4,1.4] & [-30,30] \\
        $\Phi_{4\ell{}}$, $m_{12}$  & [-1.9,1.9] & [-0.85,0.85] & [-3.7,3.7] \\
         $O_{NN}$ (binary)  & [-1.5,1.5] & [-0.75,0.75] & [-3.0,3.0] \\
          $O_{NN}$ (multi-class)  & [-1.4,1.4] & [-0.71,0.71] & [-2.7,2.7] \\
    \hline   
    \end{tabular}
    \caption{Expected 95\% confidence interval for the three Wilson coefficients given an integrated luminosity of 139~fb$^{-1}$. Results are presented for a one-dimensional fit to the $\Phi_{4\ell{}}$ distribution, a fit to double-differential yield as a function of $\Phi_{4\ell{}}$ and $m_{12}$, and fits to the $O_{NN}$ variable constructed from the neural-net outputs of the binary and multi-class models. The $O_{NN}$ variable is constructed from neural networks trained on the interference predicted by the ${\cal{O}}_{\Phi \widetilde{W}B}$ operator. 
    \label{tab:4lsummary}}
\end{table*}

\begin{figure}[t!]
\centering
    \includegraphics[width=0.45\textwidth]{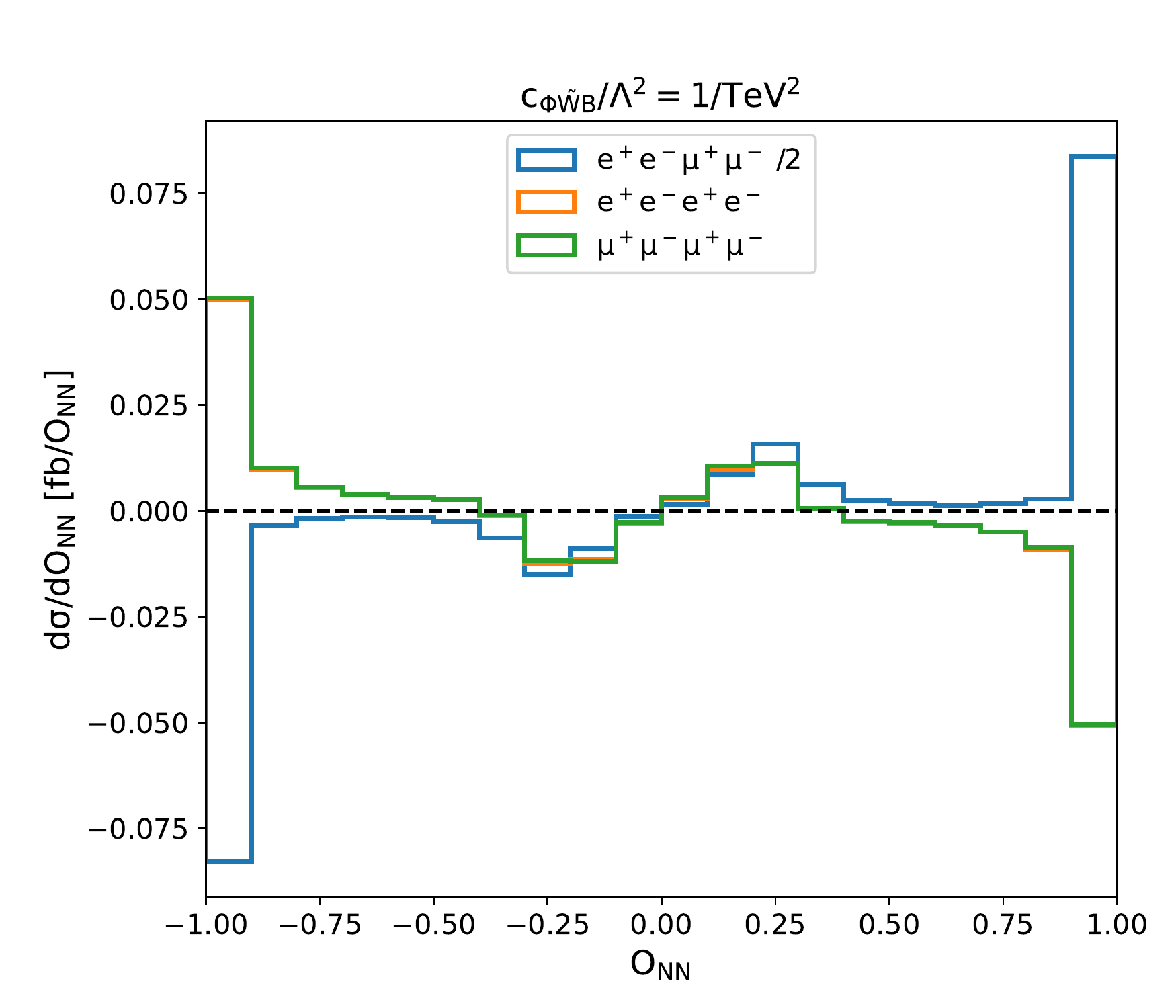}
  \caption{Differential cross section as a function of the CP-odd observable produced by a binary NN, where the NN has been trained to distinguish between the positive- and negative- interference effects produced by the ${\cal{O}}_{\Phi \widetilde{W}B}$ operator in the $h\to e^+ e^- \mu^+ \mu^-$ decay channel.}
  \label{fig:4lcomp}
\end{figure}

Finally, we discuss the analysis of  $h\to e^+ e^- e^+ e^-$ and $h\to \mu^+\mu^- \mu^+ \mu^-$. In these decay channels, there are two possible combinations of $\ell^+\ell^-$ pairs. We adopt the strategy taken in the ATLAS $4\ell{}$ analysis, whereby all possible same-flavour lepton-antilepton pairs are considered and the pair with invariant mass closest to the mass of the $Z$ boson is defined as the `first' pair (with mass $m_{12}$). The second pair is then constructed from the remaining lepton and antilepton. Figure~\ref{fig:4lcomp} shows the differential cross section as a function of the CP-odd observable produced by a binary NN, where the NN has been trained to distinguish between the positive- and negative- interference effects produced by the ${\cal{O}}_{\Phi \widetilde{W}B}$ operator in the $h\to e^+ e^- \mu^+ \mu^-$ decay channel. The model retains the capability to distinguish between the different interference contributions for the $h\to e^+ e^- e^+ e^-$ and $h\to \mu^+\mu^- \mu^+ \mu^-$ decay channels, but there are two key differences with respect to the $h\to e^+ e^- \mu^+ \mu^-$ decay channel. The first is a sign-flip in the differential cross section contribution at $O_{NN}=\pm1$; this arises due to the increase in kinematic combinations allowed for the $h\to e^+ e^- e^+ e^-$ and $h \to \mu^+ \mu^- \mu^+ \mu^-$ amplitudes leading to an inversion of the correlation of Fig.~\ref{fig:m4lphi4l}. The second feature is that the contribution at $O_{NN}\sim \pm1$ is smaller and broader, implying a poorer separation of the interference contributions. The change in sign means that the observable will need to be measured independently for each decay channel to avoid an unwanted cancellation in the asymmetry. The constraints obtained on Wilson coefficients when including the information from all three decay channels and using the CP-odd observable constructed for a binary network are found to improve by 10-20\% when compared to the constraints obtained from the $h\to e^+ e^- \mu^+ \mu^-$ decay channel alone.

\subsection{$h+2~{\text{jets}}$}
\label{sec:wbf}
We now turn to VBF as another avenue to constrain the CP structure of Higgs boson interactions. 
The phenomenology of vector boson fusion is very different to $h\to 4\ell $ because VBF is a multi-scale process, whereas the Higgs mass sets the scale for $h\to 4\ell$. 

\begin{figure}[t!]
\centering
    \includegraphics[width=0.45\textwidth]{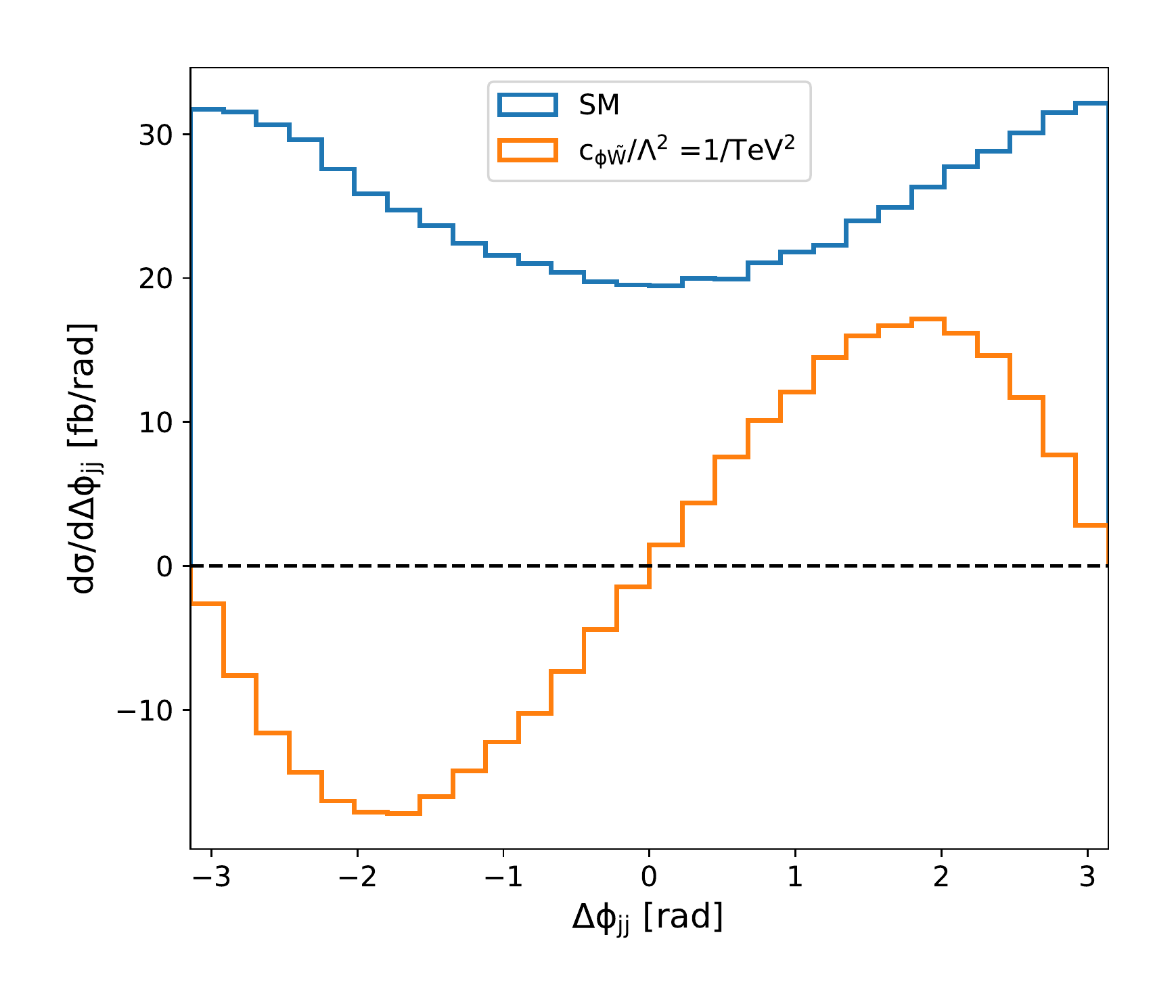}
  \caption{Differential cross section for SM and ${\cal{O}}_{\Phi \widetilde{W}}$ operator are shown as a function of the CP-odd observable $\Delta \phi_{jj}$ in VBF production of the Higgs.}
  \label{fig:2j}
\end{figure}

In Fig.~\ref{fig:2j}, we show the signed-$\Delta \phi_{jj}$ distribution of Eq.~\eqref{eq:signedphijj} for the interference contribution induced by the ${\cal{O}}_{\Phi\widetilde{W}}$ operator with $c_{\Phi \widetilde{W}}/\Lambda^2=1$~TeV$^{-2}$. This is the most important operator that affects the Higgs boson production via vector boson fusion, with the remaining electroweak operators of Eq.~\eqref{eq:ops1} playing a subdominant role (see, e.g., Refs.~\cite{Bernlochner:2018opw,Ethier:2021ydt,Dedes:2020xmo}).
The lack of sensitivity to the interference contributions induced by the ${\cal{O}}_{\Phi\widetilde{B}}$ and ${\cal{O}}_{\Phi\widetilde{W}B}$ operators arises due to the hypercharge coupling structure of the $Z$ boson interactions and the off-shellness of the $t$-channel momentum transfers, which leave a  small $Z\gamma$-interference contribution related to a small set of partonic subprocesses.

\begin{figure}[t!]
\centering
    \includegraphics[width=0.45\textwidth]{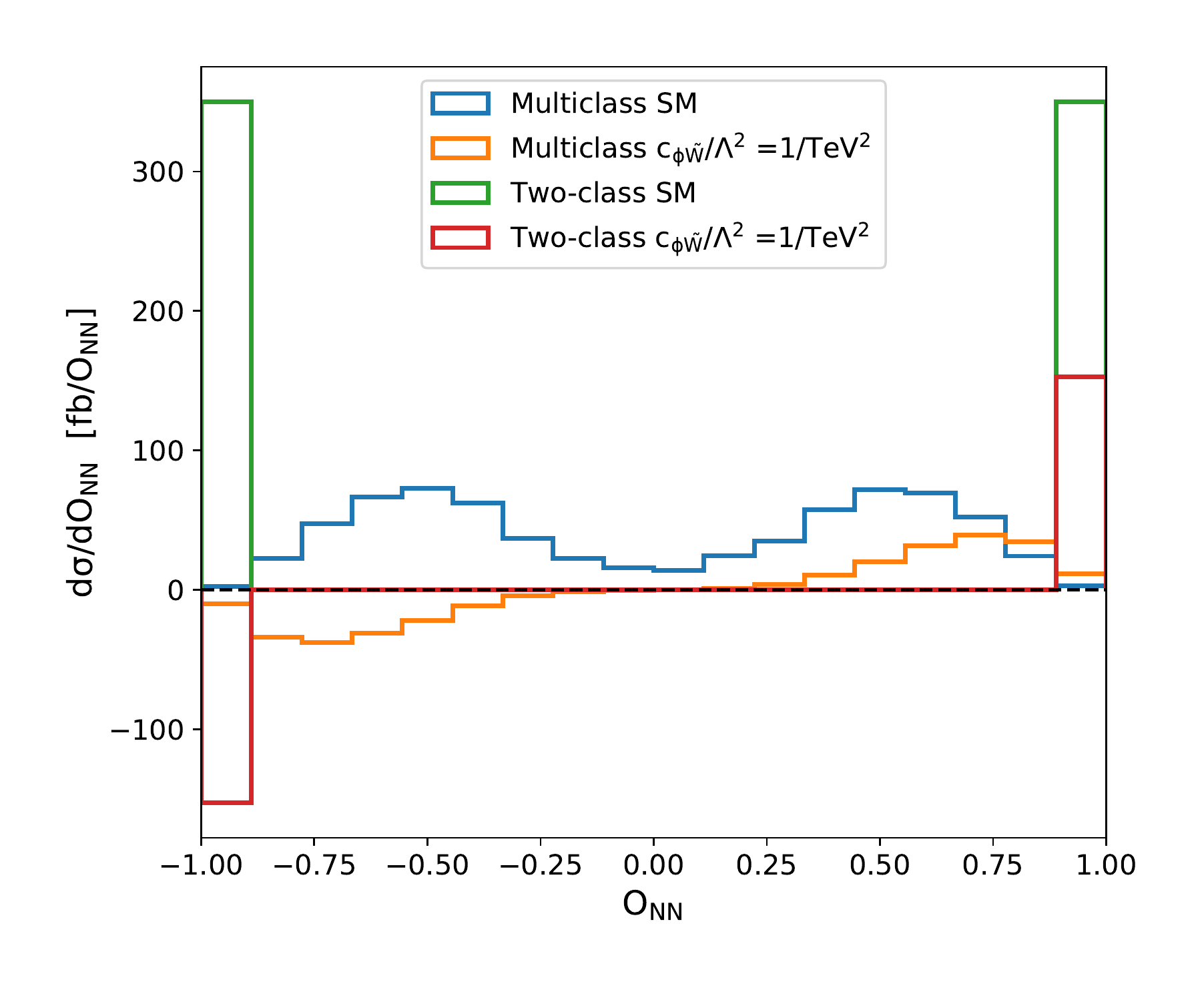}
  \caption{Differential cross section as a function of the CP-odd observable produced by a binary NN and multi-class, where the NN has been trained to distinguish between the positive- and negative- interference effects produced by the ${\cal{O}}_{\Phi \widetilde{W}}$ operator in the VBF channel. The slight asymmetry in the binary SM distribution comes from the network bias as the network is trained only on EFT.}
  \label{fig:2jcomp}
\end{figure}

The signed-$\Delta\phi_{jj}$ is found to dominate all other correlations when we perform a binary classification. Concretely, the network learns to distinguish between positive and negative interference contributions simply by projecting out the total asymmetry. This is shown in  Fig.~\ref{fig:2jcomp}, where the interference contribution and the SM contribution both populate the same bins at high $|O_{NN}|$. Any other kinematic dependence that is characteristic of a given operator is irrelevant when we only try to discriminate between positive and negative interference contributions.\footnote{We do observe the $p_{\rm T}$ enhancement that is discussed in Sec.~\ref{sec:cp}, yet the bulk of the discrimination happens at low $p_{\rm T}$ where the biggest share of the cross section is localised.}

To better exploit the underlying kinematics, we can turn to the multi-class networks, which learns the kinematic information of the SM beyond the symmetry of $\Delta \phi_{jj}$. 
This is shown in Fig.~\ref{fig:2jcomp}, where the additional information is used to discriminate between the SM contribution and the interference contributions. The SM contribution is located closer to $O_{NN}\sim 0$ than the interference contributions, implying that the multi-class network has exploited some differences in kinematics between the SM prediction and the interference prediction.

To quantify the sensitivity of each observable constructed for VBF $h+2~\rm{jets}$, the constraints on Wilson coefficients are estimated using the same likelihood setup as described in Sec.~\ref{sec:h4ell}. The expected number of events are obtained from the event generator samples after applying a normalisation factor that is defined such that the SM prediction for VBF Higgs production reproduces the number of events observed experimentally in the \textit{VBF\_1} fiducial region of Ref.~\cite{ATLAS-CONF-2021-044} (corresponding to an integrated luminosity of 139~fb$^{-1}$). The SM-only event yields are then further increased to account for background contributions from non-Higgs processes.

The constraints on the Wilson coefficients are given in Tab.~\ref{tab:vbf_limits}. For all operators considered in this work, the multi-class neural network improves the constraints when compared to the use of $\Delta \phi_{jj}$ alone. It is also clear that the kinematic information accessed via the multi-class approach is crucial for constraints on ${\cal{O}}_{\Phi\widetilde{W}}$: the binary classification does not access bin-to-bin sensitivity, which leads to a slightly decreased sensitivity compared to $\Delta\phi_{jj}$. Only the ${\cal{O}}_{\Phi\widetilde{W}}$ operator can be constrained significantly with the LHC Run-II datatset (139~fb$^{-1}$), as the constraints on ${\cal{O}}_{\Phi \widetilde{W}B}$ and ${\cal{O}}_{\Phi \widetilde{B}}$ remain too loose to by directly physically relevant. This is is in line with previous findings~\cite{Bernlochner:2018opw,Ethier:2021ydt,Dedes:2020xmo}. However, the gain in sensitivity to these operators that can be achieved by using neural networks will be important with larger datasets in the future, e.g. in LHC Run-III, at the High-Luminosity (HL) LHC, or at a Future Circular Collider.

\begin{table*}[t!]
    \centering
    \begin{tabular}{|c|c|c|c|}
    \hline 
     CP-odd observable & $c_{\Phi \widetilde{W}B} / \Lambda^2$~[TeV$^{-2}$] & $c_{\Phi \widetilde{B}} / \Lambda^2$~[TeV$^{-2}$] &  $c_{\Phi \widetilde{W}} / \Lambda^2$~[TeV$^{-2}$] \\
     \hline 
       $\Delta\phi_{jj}$  & [-21,+21] & [-149,+149] & [-0.60,+0.60] \\
     
         $O_{NN}$ (binary)  & [-11,+11] & [-43,+43] & [-0.66,+0.66] \\
          $O_{NN}$ (multi-class)  & [-10,+10] & [-36,+36] & [-0.42,+0.42] \\
    \hline   
    \end{tabular}
    \caption{Expected 95\% confidence interval for the three Wilson coefficients given an integrated luminosity of 139~fb$^{-1}$. Results are presented for a one-dimensional fit to the $\Delta\phi_{jj}$ distribution, and fits to the $O_{NN}$ variable constructed from the neural-net outputs of the binary and multi-class models. The $O_{NN}$ variable is constructed from neural networks trained on the interference predicted by the each operator separately. 
    \label{tab:vbf_limits}}
\end{table*}

\section{Summary and Conclusions}
\label{sec:conc}

In this article, we have outlined a method to directly construct CP-odd observables using the output of neural networks. The method exploits the fact that CP asymmetries arise from the interference between the SM and BSM scattering amplitudes. The neural-network is then able to optimise the separation of positive- and negative- interference contributions, using the full kinematic information that is available for a given production or decay process.

We demonstrated the performance of this method by constructing CP-odd observables for the $h\to 4\ell$ decay channel and the VBF Higgs production mechanism. Although CP-odd observables can be exploited in either channel to constrain CP-violating interactions in the Higgs sector, we have shown that the use of neural networks  can lead to large improvements in sensitivity to CP-violating effects in the Higgs sector. Specifically, we demonstrated this using dimension-six effective field theory predictions for the interference contributions. Improving the sensitivity to CP-violating effects in $h\to 4\ell$ and VBF Higgs production is particularly important for the self-consistency of the dimension-six approach~\cite{Lang:2021hnd,Araz:2020zyh}. 

In the $h\to 4\ell$ decay channel, we have shown that both binary networks and multi-class networks improve the sensitivity to CP-violating effects in the Higgs boson interactions with weak bosons, when compared to the use of traditional angular variables alone. Using the kinematic features identified by the network, we showed that the improved sensitivity derives from a sign-flip in the interference contributions when the highest-mass lepton-antilepton pair corresponds to an on-shell $Z$ boson or an off-shell $Z^*/\gamma^*$ bosons. The sign-flip in the interference term arises due to different contributions of $h\to Z\gamma$ and $h\to ZZ$ amplitudes in each region. Operators that modify the electroweak Higgs boson gauge interactions can therefore be constrained with much higher sensitivity than focusing solely on the CP-odd angular observable ($\Phi_{4\ell}$), either by using the observable constructed from the neural network output or by performing a double-differential analysis of the event yield as a function of $\Phi_{4\ell}$ and $m_{12}$. Specifically, we found that a sizeable ${\cal{O}}({10})$ improvements can be achieved using our method when compared to the use of the angular observable alone. It will be important to eventually compare the sensitivity of the CP-odd observables presented in this paper to those constructed directly from matrix-element methods~\cite{Gritsan:2020pib}.

In VBF Higgs production, the sensitivity to CP-violating effects is predominantly limited to ${\cal{O}}_{\Phi\widetilde{W}}$. We found that the angular observable $\Delta\phi_{jj}$ drives this sensitivity. However, the multi-class network outperforms the angular observable, as it accesses the full kinematic information of each class and tensions the interference contribution against the SM contribution. We note that the application of multi-class machine learning improves the sensitivity to the phenomenologically less significant EFT operators by at least a factor of two, which is equivalent to quadrupling the integrated luminosity of the dataset. Our neural-net-based method will therefore allow these operators to be scrutinised in detail at the HL-LHC.

We note that the construction of CP-odd observables using neural networks can be generalised to many other processes that probe CP-violation at the LHC, including other Higgs boson production and decay channels, as well as searches for CP-violating effects in the weak-boson self-interactions. Although we have focused on a SMEFT analysis, the techniques presented in this work directly generalise to light propagating BSM degrees of freedom that could induce CP violation (this could be captured via retaining the full mass dependence of the Wilson coefficients on the BSM particles). In a similar spirit, absorptive parts of SM amplitudes~\cite{Frederix:2014cba} could be analysed via the introduced classification, thus providing a novel angle on validating QCD predictions. While more traditional approaches, (i.e. measuring angular observables in $h\to 4\ell$ and $pp\to h+2~\text{jets}$) remain important tools for the clarification of the CP-structure of the Higgs sector, their generalisation to more comprehensive BSM classifiers also taking into account additional correlations will enhance the sensitivity of analyses during LHC Run-III, the HL-LHC, and at a potential Future Circular Collider.

\bigskip
\noindent{\bf{Acknowledgements}} ---
We thank Florian Bernlochner and Andrei Gritsan for useful conversations and insightful comments on early versions of this work. A.B. and C.E. are supported by the STFC under grant ST/T000945/1. C.E. is supported by the Leverhulme Trust under grant RPG-2021-031 and the IPPP Associateship Scheme. A.D.P is supported by the Royal Society and STFC under grants UF160396 and ST/000925/1. R.H. and A.D.P are supported by the Leverhulme Trust under grant RPG-2020-004.
The data supporting the findings reported in this paper are openly available from the figshare repository \cite{dataonline}.


\end{document}